\documentclass[journal,12pt,draftclsnofoot,onecolumn]{IEEEtran}
\IEEEoverridecommandlockouts
\usepackage[utf8]{inputenc}
\usepackage[T1]{fontenc}
\usepackage{url}
\usepackage{ifthen}
\usepackage{cite}
\usepackage{amsmath} 

\usepackage{amsthm}
\usepackage{amsfonts}
\usepackage{mathrsfs}
\usepackage{multirow}
\usepackage{citesort}
\usepackage{subfigure}
\usepackage{bm}
\usepackage{algorithm}
\usepackage{graphicx}
\usepackage{epstopdf}
\usepackage{amssymb}
\usepackage{diagbox}
\usepackage{makecell}
\usepackage{caption}
\usepackage{xhfill}
\allowdisplaybreaks[4]
\newcommand{\xfilll}[2][1ex]{%
	\dimen0=#2\advance\dimen0 by #1%
	\leaders\hrule height \dimen0 depth -#1\hfill%
}

\begin{document}
\title{Transmission of Bernoulli Sources Using Convolutional LDGM Codes\\
\thanks{Corresponding author: Xiao Ma. This work was supported by the National Key R\&D Program of China~(Grant No.~2021YFA1000500) and the NSF of China~(No.~61971454). Part of this work is accepted by 2022 International Symposium on Information Theory. }
}

\author{
  \IEEEauthorblockN{  Yixin Wang, Tingting Zhu and Xiao Ma,~\emph{Member}, IEEE}\\
 \thanks{ The authors are with the  Guangdong Key Laboratory of Information Security Technology, School of Computer Science and Engineering, Sun Yat-sen University,  Guangzhou 510006, China~(e-mail:
 	wangyx58@mail2.sysu.edu.cn, zhutt@mail2.sysu.edu.cn, maxiao@mail.sysu.edu.cn)}
}

\maketitle

\begin{abstract}
We propose in this paper to exploit convolutional low density generator matrix~(LDGM) codes for transmission of Bernoulli sources over binary-input output-symmetric~(BIOS)  channels. To this end, we present a new framework to prove the coding theorems for linear codes, which unifies the channel coding theorem, the source coding theorem and the joint source-channel coding~(JSCC) theorem. In the presented framework, the systematic bits and the corresponding parity-check bits play different roles. Precisely, the noisy systematic bits are used to limit the list size of typical codewords, while the noisy parity-check bits are used to select from the list the maximum likelihood codeword. This new framework for linear codes allows that the systematic bits and the parity-check bits are transmitted in different ways and over different channels. With this framework, we prove that the Bernoulli generator matrix codes~(BGMCs) are capacity-achieving over BIOS channels, entropy-achieving for Bernoulli sources, and also system-capacity-achieving for JSCC applications. A lower bound on the bit-error rate~(BER) is derived for linear codes, which can be used to predict the error floors and hence serves as a simple tool to design the JSCC system. Numerical results show that the convolutional LDGM codes perform well in the waterfall region and match well with the derived error floors, which can be lowered down if required by simply increasing the encoding memory.
\end{abstract}

\section{Introduction}
\par Shannon's separation theorem~\cite{Shannon1948} states that, with unbounded block length~(delay) and complexity, the channel coding and the source coding can be separated without loss of optimality. Evidently, for most applications in the non-asymptotic regime with limited delay and complexity, a joint source-channel coding~(JSCC) can be more attractive.  A commonly accepted JSCC scheme consists of two component codes, one as the source code for compression and the other as the channel code for error correction, where the redundancy left by the source encoder can be exploited by the channel decoder to improve the system performance.  Typical constructions include the double low-density parity-check (D-LDPC) JSCC and the double polar~(D-Polar) JSCC. The D-LDPC JSCC was
first proposed in~\cite{Maria2009,verdu2010} and later evolved as double protograph LDPC~(DP-LDPC) joint source-channel codes~(JSCCs)~\cite{He2012} and double spatially coupled LDPC~(D-SCLDPC) JSCCs~\cite{Golmohammadi2021} by taking either propotograph LDPC codes or SC-LDPC codes as the component codes. The D-Polar JSCC was investigated in~\cite{Dong2021}, where a turbo-like belief propagation~(BP) decoder was proposed. In any case, the two component codes need to be designed  jointly, which usually involves a complicated optimization algorithm such as the extended curve-fitting algorithm~\cite{Chen2018} and  multi-objective differential evolution algorithm~\cite{Liu2020}. Since a concatenation of two linear codes is still linear, it is reasonable to develop a single linear code to achieve JSCC, as an attempt in~\cite{Lau2021}. In this paper, we will first prove in theory that one single linear code~(even with low density generator matrix~(LDGM)) is sufficient to achieve the JSCC limit and then provide concrete constructions  for illustration. The most distinguished feature of the proposed system is that it does not require complicated optimization.   
\par Linear codes play an important role in both the channel coding theory~\cite{Gallager1968} and the source coding theory~\cite{Csizar2011}. In\cite{Elias1955Coding}, it was proved that the totally random linear code ensemble can achieve the capacity of binary-input output-symmetric (BIOS) channels. The same theorem was proved in~\cite{Gallager1968} by deriving the error exponent. Systematic linear codes have the information bits in the codewords, which can benefit the encoding and decoding procedure compared with non-systematic linear codes. Of the same codeword length, systematic linear codes can have less operating steps in the coding procedure. More importantly, using systematic instead of non-systematic linear codes allows the decoder to obtain the decoded bits directly from the received sequences.
However, most existing coding theorems are proved for non-systematic codes and direct proofs are rarely found for systematic codes.

\par In this paper, we propose to transmit reliably a Bernoulli source over a binary-input output symmetric~(BIOS) channel by employing a single linear code, which integrates the source coding and the channel coding as a seamless system.  To this end, we propose a new framework to prove the coding theorems for linear codes, which unifies the channel coding theorem, the source coding theorem and the JSCC theorem. In the presented framework, the systematic bits and the corresponding parity-check bits play different roles. Precisely, the noisy systematic bits are used to limit the list size of typical codewords, while the noisy parity-check bits are used to select from the list the maximum likelihood codeword. This new framework for linear codes allows that the systematic bits and the parity-check bits are transmitted in different ways and over different channels. With this framework, we prove that the Bernoulli generator matrix codes~(BGMCs) are capacity-achieving over BIOS channels, entropy-achieving for Bernoulli sources, and  also system-capacity-achieving for JSCC applications.  A lower bound on the bit-error rate~(BER) is also derived for linear codes, which can be used to predict the error floors and hence serves as a simple tool to design the JSCC system. 
For the simulations in JSCC, we consider a special class of linear codes called convolutional LDGM codes. Numerical results show that the convolutional LDGM codes are flexible to construct and have predictable error floors. The convolutional LDGM codes also perform well in the waterfall region with about one dB away from the system capacity.

\par The rest of the paper is organized as follows. In Section~\ref{sec2}, we describe the new framework and give an overall coding theorem for the framework. In Section~\ref{sec3}, we prove the coding theorem for the framework  with BGMCs. In Section~\ref{sec4}, the systematic capacity for transmitting Bernoulli sources is derived from the coding theorem and the performance lower bound for linear codes in JSCC scheme is also derived. We use the convolutional LDGM codes in the  framework for transmitting Bernoulli sources.  Numerical results for various parameter settings show that the convolutional LDGM codes have flexible construction, good performance in the waterfall region and match well with the lower bound in the  error floor region. Finally, some concluding remarks are given in Section~\ref{sec5}.
\par In this paper, a random variable is denoted by an upper-case letter, say $X$, whose realization is denoted by the corresponding lowercase letter $x\in\mathcal{X}$. We use $P_{X}(x)$, $x\in \mathcal{X}$ to represent the probability mass~(or density) function of a random discrete~(or continuous) variable. For a vector of length $m$, we represent it as $\bm x=(x_0,x_1,\cdots,x_{m-1})$. We also use $\bm x^m$ to emphasize the length of $\bm x$. We denote by $\mathbb{F}_2=\{0,1\}$ the binary field.
\section{System model and Problem Statement}
\label{sec2}

\subsection{Systematic Linear Codes}

\par
\begin{figure}[t]
	\centering
	\includegraphics[width=0.5\textwidth]{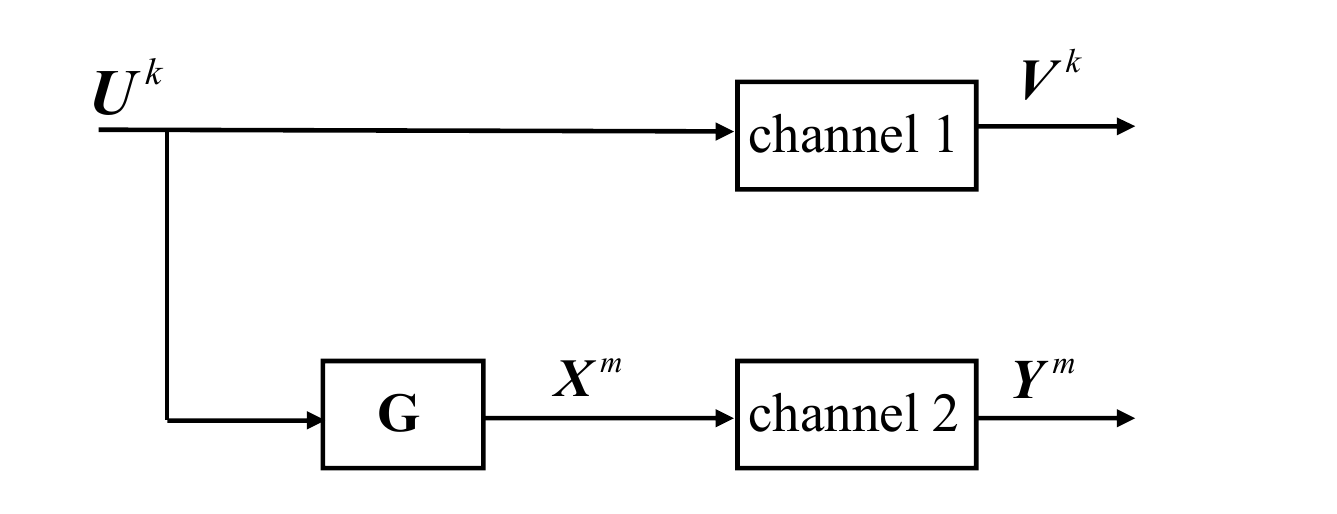}
	\caption{A  system model with linear coding. For the source coding, $\bm V^k$ is punctured and the code rate is $m/k$, while for the channel coding, the code rate is defined as usual $k/(k+m)$.}
	\label{systemm model}
\end{figure}

 We consider a system model that is depicted in Fig.~\ref{systemm model}, where $\bm U^{k}\in \mathbb{F}_2^{k}$ is referred to as the information bits to be transmitted, ${\rm{\bf G}}$ is a binary matrix of size $k\times m$ and $\bm X^m=\bm U^k{\rm{\bf G}}\in \mathbb{F}_{2}^{m}$ is referred to as parity-check bits corresponding to $\bm U^k$. The two channels, channel~$1$ and channel~$2$ are independent and can be different. For the channel coding, channel~$1$ and channel $2$ are typically the same. For the source coding, $\bm V^k$ is not required and channel $1$ can be viewed as an erasure channel with erasure probability one while channel~$2$ is a noiseless channel. The model can also be adapted to more general cases such as the JSCC. Even more generally, both channel~$1$ and channel~$2$ can be integrated with high-order modulations and demodulations. The task of the receiver is to recover $\bm U^k$ from $(\bm V^k,\bm Y^m)$.
\par In this paper, we focus on Bernoulli sources and BIOS memoryless channels. A Bernoulli source~(also referred to as a Bernoulli sequence or a Bernoulli process) is a sequence of independent and identically distributed~(i.i.d.) binary random variables. A  BIOS  memoryless channel is characterized by the input  $x \in \mathcal{X}=\mathbb{F}_{2}$, an output set $\mathcal{Y}$~(discrete or continuous), and a conditional probability mass (or density) function~\footnote{If the context is clear, we may omit the subscript of the probability mass (or density) function.} $\{P_{Y|X}(y|x), x\in\mathbb{F}_2, y\in\mathcal{Y}\}$ which satisfies the symmetric condition that $P_{Y|X}(y|1)=P_{Y|X}(\pi(y)|0)$ for some mapping $\pi~: \mathcal{Y}\rightarrow\mathcal{Y}$ with $\pi^{-1}(\pi(y))=y$. The channel is said to be memoryless if $P_{\bm{Y}|\bm{X}}(\bm y|\bm x)=\prod\limits_{i=0}^{n-1}P_{Y|X}(y_i|x_i)$. For a Bernoulli source $\bm U$, we can define the conditional entropy $H(U|V)$. For a Bernoulli input $\bm X$, we can define the mutual information $I(X; Y)$. Obviously, the parity-check bits are necessary only when $H(U|V) > 0$.

\subsection{A General Coding Theorem for Linear Codes}
\emph{Theorem 1:} Let $\bm U$ be a Bernoulli source and ${\rm {\bf G}}$ be a totally random matrix, whose elements are generated independently and identically according to the Bernoulli distribution with success probability $\rho=1/2$. For any fixed nonnegative number $r<I(X;Y)/H(U|V)$, there exists a matrix ${\rm {\bf G}}$ of sufficiently large size and a decoding algorithm such that $k/m \geq r$ and the frame-error rate (FER) ${\rm Pr}\{\hat{\bm U}\neq \bm U\}$ is arbitrarily small.
 \par \emph{Outline  Proof of Theorem 1:} Let $\epsilon$ be an arbitrarily small but fixed positive number. Upon receiving $(\bm v,\bm y)$, we conduct the following two-step decoding. First, list all sequences $\bm {\tilde u}$ such that $(\bm {\tilde u},\bm v)$ are jointly typical. Second, find from the list a sequence $\hat{\bm u}$ such that $P(\bm y|\hat{\bm u}{\rm {\bf G}})$ is maximized.
 \par There are two types of errors. One is the case when $\bm u$ is not in the list and the other is the case when $\bm u$ is in the list but is not the most likely one. The former error event can have arbitrarily small probability since $(\bm U,\bm V)$ are jointly typical with  arbitrarily high probability for sufficiently large $k$. The latter error event can also have arbitrarily small probability as long as the list size (which is upper bounded by $\exp[kH(U|V)+2\epsilon]$ is less than the capacity of channel~$2$. This can be fulfilled by choosing sufficiently  large $k$ and $m$ such that $k/m \geq r$ but $k/m < I(X;Y) / H(U|V)$ since $r < I(X;Y) / H(U|V)$ from the assumption of the theorem.

\hfill $\square$
\par \emph{Remarks:} The above theorem states that the ratio $k/m$ can be arbitrarily close to  $I(X;Y)/H(U|V)$. Actually, we can prove that, for ${\rm Pr}\{\hat{\bm U}\neq \bm U\}\rightarrow0$,

\begin{equation}
	\varlimsup\limits_{\substack{
			k\rightarrow\infty\\m\rightarrow\infty}}\frac{k}{m}=\frac{C}{H(U|V)}\text{,}
\end{equation}
where $C$ is the channel capacity of channel~$2$.

\begin{itemize}
	\item For the Bernoulli source with $P(0)=P(1)=1/2$ and the same channel~$1$ and channel~$2$, the condition $k/m<I(X;Y)/H(U|V)$ is equivalent to the conventional condition $k/(m+k)<I(X;Y)$, where $k/(k+m)$ is the code rate for channel coding. This can be verified by noting that $I(X;Y)=1-H(X|Y)$ and $H(U|V)=H(X|Y)$.
	\item For the source coding, channel~$1$ is an erasure channel and channel~$2$ is noiseless. Hence, the condition  $k/m<I(X;Y)/H(U|V)$ is equivalent to the conventional condition $m/k>H(U)$, where $m/k$ is the code rate for the source coding. This can be verified by noting that $H(U|V)=H(U)$ and $I(X;Y)=1$.
	\item From the proof, we see that the systematic bits and the parity-check bits play different roles. Receiving noisy systematic bits provides us a list of the source output, while receiving noisy parity-check bits helps us to select the correct one from the list.
\end{itemize}
\par It is well-known that the linear codes defined by general matrices ${\rm {\bf G}}$  have no efficient decoding algorithms for large $k$ and $m$. Hence, we consider the following LDGM code ensemble~\cite{Ma2016Coding}, which is referred to as Bernoulli generator matrix code (BGMC).

\par \emph{BGMC ensemble:} A BGMC transforms $\bm u\in \mathbb{F}_{2}^{k}$ into $(\bm u,\bm x)$ by $\bm x=\bm u{\rm {\bf G}}\in \mathbb{F}_{2}^{m}$, where {\rm {\bf G}} is a random matrix of size $k \times m$ with each element $G_{i,j}~(0\leq i\leq k-1,~0\leq j \leq m-1)$ being generated independently according to the Bernoulli distribution with success probability ${\rm Pr}\{G_{i,j}=1\}=\rho \leq 1/2$. For $\rho \ll 1/2$, the BGMC ensemble is a class of LDGM codes.

\par It has been proved that, in terms of BER, the BGMC ensemble is capacity-achieving for BIOS memoryless channels~\cite{Ma2016Coding,2020SCLDGM} and entropy-achieving for Bernoulli sources~\cite{Zhu2021Compression}. In this paper,  we show that Theorem $1$ also holds
in terms of FER even for $\rho<1/2$. The proof is definitely applicable to the case  $\rho=1/2$, indicating that the outline proof of Theorem~1 will be detailed in the next section.
\section{Coding Theorem for BGMCs}
\label{sec3}
\subsection{Partial Mutual Information}
Let $P(1)=p$ and $P(0)=1-p$~be an input distribution of a BIOS memoryless channel. The mutual information between the input and the output is given by

\begin{equation}
	I(p)=(1-p)I_0(p)+pI_1(p)\text{,}
\end{equation}
where
\begin{equation}
	I_0(p)=\sum\limits_{y\in\mathcal{Y}} P(y|0)\log\frac{P(y|0)}{P(y)}\text{,}
	\label{I_{0}}
\end{equation}
\begin{equation}
	I_1(p)=\sum\limits_{y\in\mathcal{Y}} P(y|1)\log\frac{P(y|1)}{P(y)}\text{,}
\end{equation}
and $P(y)=(1-p)P(y|0)+pP(y|1)$. We define $I_0(p)$ $\big({\rm or}~I_1(p)\big)$  as \emph{partial mutual information}. For a BIOS memoryless channel, we have ${\rm max}_{0\leq p \leq 1}I(p)=I(1/2)=I_0(1/2)=I_1(1/2)$, which is the channel capacity. Notice that $I_0(p) > 0$ for $0<p<1$ as long as ${\rm Pr}\{y |P(y|0) \neq P(y|1)\} > 0$. This is a natural assumption in this paper.

\par \emph{Lemma 1:}  The partial mutual information $I_0(p)$ is continuous, differentiable and strictly increasing from $I_0(0)=0$ to the capacity $I_0(1/2)$.
\par \emph{Proof:} It can be easily seen that the partial mutual information is continuous and differentiable for $0\leq p\leq 1/2$. By carrying out the differentiation, we can verify that partial mutual information is strictly increasing from $I_0(0)=0$ to the capacity $I_0(1/2)$.  \hfill $\square$
\par To prove the main theorem, we also need the following lemmas.
\par\emph{Lemma 2:} 	For the BGMC ensemble, the parity-check vector corresponding to an information vector with weight $\omega$ is a Bernoulli sequence with success probability
\begin{equation}
	\rho_{\omega}\triangleq{\rm{Pr}}\{X_j=1|W_H(\bm U)=\omega\}=\frac{1-(1-2\rho)^{\omega}}{2}\text{,}
\end{equation}
where $W_H(\cdot)$  is the Hamming weight function. Furthermore, for any given positive integer $T\leq k$,
\begin{equation}
	\begin{aligned}
		P(\bm x|\bm u)&\triangleq {\rm Pr}\{\bm X=\bm x|U=\bm u \}\\
		&\leq P(\bm 0|\bm u)\leq(1-\rho_{T})^{m}\text{,}
	\end{aligned}
\end{equation}
for all $\bm u\in \mathbb{F}_{2}^{k}$ with $W_H(\bm u)\geq T$ and $\bm x\in \mathbb{F}_{2}^{m}$.

\par \emph{Proof:} See the proof in\cite[Lemma 1]{2020SCLDGM} and it is omitted here.\hfill $\square$
\par\emph{Remark:} Lemma 2 states that, for information vector $\bm u$ with high weight, the corresponding parity check vector is convergent in distribution to a Bernoulli process with success probability $1/2$, since $\rho_{\omega}\rightarrow 1/2$ as $\omega\rightarrow \infty$.
\par \emph{Lemma 3:} Given a sequence $\bm v$ of length $k$, for any $\epsilon>0$, define $A_{\epsilon}^{(k)}(\bm U|\bm v)$  the set of $\bm u$ sequences which are jointly $\epsilon$-typical  with $\bm v$. If $\bm v$ is typical,  then for sufficiently large $k$, the cardinality of $A_{\epsilon}^{(k)}(\bm U|\bm v)$ can be upper bounded by
\begin{equation}
	\big|A_{\epsilon}^{(k)}(\bm U|\bm v)\big|\leq \exp\Big[k(H(U|V)+2\epsilon)\Big].
\end{equation}
\par \emph{Proof: } See \cite[Theorem 15.2.2]{Cover2006} and it is omitted here.\hfill $\square$
\subsection{Partial Error Exponent}
\par The error exponent was derived to prove the channel coding theorem by assuming that all codewords are randomly generated according to an identical distribution, where the pair-wise independence between codewords is sufficient\cite{Gallager1968}. In this paper, we derive the partial error exponent for BIOS channels by assuming that the codeword $\bm 0$ is transmitted. The derivation suggests that even the pair-wise independence is not required.
\par \emph{Lemma 4:} Suppose that the codeword $\bm 0\in\mathbb{F}_2^{n}$ is transmitted over a BIOS channel. Let $\mathscr{L}=\{\bm x_1,\bm x_2,\cdots,\bm x_M\}$ be a random list, where $\bm x_i\in\mathbb{F}_2^n$ is a segment of a Bernoulli process with success probability $p$. Then the probability that there exists some $i$ such that $\bm x_i$ is more likely than $\bm 0$, denoted by ${\rm Pr}\{{\rm error}|\bm 0\}$, can be upper bounded by

\begin{equation}
	{\rm Pr}\{{\rm error}|\bm 0\}\leq \exp\bigg(-nE(p,R)\bigg)\text{,}
\end{equation}
where
\begin{equation}
	R=\frac{1}{n}\log M,
\end{equation}
\begin{equation}
	E(p,R)=\max\limits_{0\leq\gamma\leq 1}(E_0(\gamma,p)-\gamma R)\text{,}
\end{equation}
and
\begin{equation}
	\begin{aligned}
		E_0(\gamma,p)=-\log\Bigg\{\sum\limits_{ y\in \mathcal{Y}}P( y| 0)&^{\frac{1}{1+\gamma}}\Big[(1-p)(P( y| 0))^{\frac{1}{1+\gamma}}+p(P( y| 1))^{\frac{1}{1+\gamma}}\Big]^\gamma\Bigg\}\text{.}
	\end{aligned}	
	\label{E independent}
\end{equation}
Furthermore, $E(p,R)>0$ if $0<R<I_0(p)$.

\par \emph{Proof: }
Denote by $E_i$ the event that $\bm x_i$ is more likely than $\bm 0$ given a received sequence $\bm y$. For the decoding error, we have
\begin{equation}\small
	\begin{aligned}
		{\rm Pr}[{\rm error}|\bm 0]&=\sum\limits_{\bm y\in \mathcal{Y}^{n}}P(\bm y|\bm 0)\cdot {\rm {Pr}}\Big\{{\bigcup\limits_{i=1}^ME_i\Big\}}\\
		&\leq  M^\gamma\sum\limits_{\bm y\in \mathcal{Y}^{n}}P(\bm y|\bm 0)\bigg({\rm {Pr}}\{P(\bm y|\bm x)\geq P(\bm y|\bm 0)\}\bigg)^\gamma\text{,}
		\end{aligned}
	\label{P}	
\end{equation}
for any given $0\leq \gamma\leq 1$. From Markov inequality, for  $s=1/(1+\gamma)$ and a given received vector $\bm y$, the probability of a vector $\bm $ being more likely than $\bm 0$ is upper bounded by
\begin{equation}
	\begin{aligned}
		{\rm Pr}\{P(\bm y|\bm x)\geq P(\bm y|\bm 0)\}&\leq  \frac{{  \rm{\bf E}}[(P(\bm y|\bm x))^{s} ]}{(P(\bm y|\bm 0))^{s}}\\
		&=\sum\limits_{\bm  x} P(\bm x)\frac{(P(\bm y|\bm x))^{s}}{(P(\bm y|\bm 0))^{s}}\text{.}
	\end{aligned}
\label{Markov}
\end{equation}

Substituting this bound into \eqref{P}, we have
\begin{equation}\small
	\begin{aligned}
	&{\rm Pr}\{{\rm error}|\bm 0\}\leq M^\gamma \sum\limits_{\bm y\in \mathcal{Y}^{n}}P(\bm y|\bm 0) \Big[\sum\limits_{\bm  x} P(\bm x)\frac{(P(\bm y|\bm x)^{s}}{(P(\bm y|\bm 0))^{s}}\Big]^\gamma\\
	&\overset{(*)}{=}M^\gamma\prod\limits_{i=0}^{n-1}\Bigg\{\sum\limits_{ y_i\in \mathcal{Y}}P(y_i|0)^{1-s\gamma}\Big[\sum\limits_{x_i\in \mathbb{F}_2} P(x_i)(P(y_i|x_i))^{s}\Big]^\gamma\Bigg\}\\
	&\overset{(**)}{\leq}\exp\bigg(-nE(p,R)\bigg)\text{,}
	\end{aligned}
\end{equation}
where the equality $(*)$ follows from the memoryless channel assumption and the inequality $(**)$ follows by recalling  that $s=1/(1+\gamma)$ and denoting
\begin{equation}
	E(p,R)=\max\limits_{0\leq\gamma\leq 1}(E_0(\gamma,p)-\gamma R)\text{,}
\end{equation}
and
\begin{equation}
	\begin{aligned}
		E_0(\gamma,p)=-\log\Bigg\{\sum\limits_{ y\in \mathcal{Y}}P( y| 0)^{\frac{1}{1+\gamma}}&\Big[(1-p)(P( y| 0))^{\frac{1}{1+\gamma}}+p(P( y| 1))^{\frac{1}{1+\gamma}}\Big]^\gamma\Bigg\} \text{.}
	\end{aligned}	
\end{equation}

Considering $E_0(\gamma,p)-\gamma R$ for a given $p$, we have $E_0(0,p)-0\cdot R=0$ and
\begin{equation}
		\frac{\partial E_0(\gamma,p)}{\partial \gamma}-R\Bigg|_{\gamma=0}=I_{0}(p)-R\text{.}
\end{equation}
Hence, $E(p, R) > 0 $ if $ R < I_0(p)$.
\hfill $\square$
\par \emph{Remark:} From the above proof, we see that the members in the list need to have the same distribution but the condition of pair-wise independence is not necessary, which is distinguished from the proof~\cite[Chapter 5]{Gallager1968}.

\subsection{Coding Theorem for BGMCs}
\label{AchievableForBGMCs}

\par \emph{Theorem 2:} Consider the BGMC ensemble and let $\bm U$ be a Bernoulli source. For any fixed nonnegative number $r<I(X;Y)/H(U|V)$, there exists a matrix ${\rm {\bf G}}$ of sufficiently large size and a decoding algorithm such that $k/m \geq r$ and the FER  is arbitrarily small.

\par \emph{Proof:}   Suppose that $(\bm u^k,\bm u^k{\rm {\bf G}})$ is transmitted. Let $\epsilon>0$ be an arbitrarily small number. Upon receiving $(\bm v^k,\bm y^m)$, we use the following two-step decoding. First, list all sequences $\bm {\tilde u}$ such that $(\bm {\tilde u},\bm v)$ are jointly typical. Second, find from the list a sequence $\hat{\bm u}$ such that $P(\bm y|\hat{\bm u}{\rm {\bf G}})$ is maximized.
\par There are two types of errors. One is when $(\bm u^k,\bm v^k)$ are not jointly $\epsilon-$typical and hence $\bm u^k$ is not in the list. This type of errors, from \cite[Chapter 7]{Cover2006}, can have arbitrarily small probability as long as $k$ is sufficiently large.
\par The other case is that $\bm u^k$ is in the list but is not the most likely one. In this case, denote the list as $\tilde{\mathscr{L}}=\{\tilde{\bm u}_0=\bm u^k,\tilde{\bm u}_1,\tilde{\bm u}_2,\cdots,\tilde{\bm u}_M\}$. From Lemma~3, we have $M\leq \exp\big[k(H(U|V)+2\epsilon)\big]$. Given the received sequence $\bm y$ and  the list  $\tilde{\mathscr{L}}$, the decoding output $\hat{\bm U}$ is a random sequence over the code ensemble due to  the randomness of $\rm{\bf G}$. Denote by $E_{\bm u,i}$ the event that $\tilde{\bm u}_i{\rm{\bf G}}$ is more likely than $\bm u^k{\rm{\bf G}}$ given a received sequence $\bm y$.  We have
\begin{equation}
	\begin{aligned}
		{\rm{Pr}}\{{\rm error}|\bm u^k\}
		&\leq\epsilon+\sum\limits_{\bm x \in \mathbb{F}_2^m }\sum\limits_{\bm y \in \mathcal{Y}^{m}} \Pr \{ \bm u^k{\rm {\bf G}} = \bm x  \} \cdot P(\bm y|\bm x)\cdot {\rm {Pr}}\Big\{{\bigcup\limits_{i=1}^ME_{\bm u,i}\Big\}}.
	\end{aligned}	
\end{equation}

Given $\bm y\in \mathcal{Y}^{m}$, $\bm x\in\mathbb{F}_2^{m}$, we define
\begin{equation}
	\pi^{x_i}(y_i)=
	\begin{cases}
		y_{i}& {\rm if~} x_i=0\\
		\pi(y_i)& {\rm if~}x_i=1
	\end{cases}.
\end{equation}
Then we have $P(\bm y|\bm x)=P(\pi^{\bm x}(\bm y)|\bm 0)$ and $P(\bm y|\tilde{\bm u}{\rm {\bf G}})\geq P(\bm y|{\bm u^k}{\rm {\bf G}} )$ is equivalent to $P(\pi^{\bm x}(\bm y)|(\bm u^k+\tilde{\bm u}){\rm {\bf G}})\geq P(\pi^{\bm x}(\bm y)|\bm 0)$. Therefore, the performance can be analyzed by assuming that $\bm 0$ is transmitted and competed with a list of codewords $\bm u_i{\rm {\bf G}}$ at the decoder, where $\bm u_i=\bm u^k+\tilde{\bm u}_i$ for $1\leq i\leq M$. With this equivalence, we have
\begin{equation}
	\begin{aligned}
		&\sum\limits_{\bm x \in \mathbb{F}_2^m } \Pr \{ \bm u^k{\rm {\bf G}} = \bm x  \} \sum\limits_{\bm y \in \mathcal{Y}^{m}}P(\bm y|\bm x)\cdot {\rm {Pr}}\Big\{{\bigcup\limits_{i=1}^ME_{\bm u,i}\Big\}}\\
		&=\sum\limits_{\bm y\in \mathcal{Y}^{m}}P(\bm y|\bm 0)\cdot {\rm {Pr}}\Big\{{\bigcup\limits_{i=1}^ME_i\Big\}},
		\label{FER}
	\end{aligned}	
\end{equation}
where $E_i$ is the event that $\bm u_i{\rm{\bf G}}$ is more likely than $\bm 0$ given a received sequence $\bm y$.
\par  We partion the list $\mathscr{L}=\{\bm 0,\bm u_1,\bm u_2,\cdots,\bm u_M\}$ according to the weight of  ${\bm u}_i~(0\leq i\leq M)$ and denote by $\mathscr{L}_\omega$ all the sequences of $\bm u_i\in\mathscr{L}$ with $W_H(\bm u_i)=\omega$. Thus, we have
\begin{equation}
    \setlength{\abovedisplayskip}{1.5pt}
	\setlength{\belowdisplayskip}{1.5pt}
	\mathscr{L}=\bigcup\limits_{\omega=0}^{k}\mathscr{L}_{\omega}\text{,}
\end{equation}
and
\begin{equation}
    \setlength{\abovedisplayskip}{1.5pt}
	\setlength{\belowdisplayskip}{1.5pt}
	\big|\mathscr{L}_{\omega}\big|\leq\binom{k}{\omega}\text{.}
	\label{boundnumber}
\end{equation}
\par For any positive integer  $T<k$, the error event can be split into two sub-events depending on whether $W_H(\bm U)\geq T$ or not. Thus, we have

\begin{equation}\small
	\begin{aligned}
		{\rm {Pr}}\Big\{{\bigcup\limits_{i=1}^ME_i\Big\}}
		&\leq\sum\limits_{\omega=1}^{T-1}\Bigg(\sum\limits_{\bm u_i: W_H(\bm u_i)=\omega}{\rm Pr}\{P(\bm y|\bm u_i{\rm {\bf G}})\geq P(\bm y|\bm 0)\}\Bigg)^{\gamma}+\Bigg(\sum\limits_{\bm u_i: W_H(\bm u_i)\geq T}{\rm Pr}\{P(\bm y|\bm u_i{\rm {\bf G}})\geq P(\bm y|\bm 0)\}\Bigg)^{\gamma}\text{,}
	\end{aligned}
\label{bound1}
\end{equation}
for any $ 0\leq\gamma\leq 1$.

For $\omega \geq1$, we define FER($\omega$) as
\begin{equation}\small
	\begin{aligned}
		{\rm FER}(\omega)&=\sum\limits_{\bm y\in \mathcal{Y}^m}P(\bm y|\bm 0)\cdot\Bigg(\sum\limits_{\bm u_i: W_H(\bm u_i)=\omega}{\rm Pr}\{P(\bm y|\bm u_i{\rm {\bf G}})\geq P(\bm y|\bm 0)\}\Bigg)^{\gamma}\text{.}
	\end{aligned}
\end{equation}

We know from Lemma 2 that, for any given $\bm u_i$ with $W_H(\bm u_i) = \omega$, the parity-check vector $\bm x$ is a segment of Bernoulli process with success probability $\rho_{\omega}$.
Hence, the conditional probability mass function $P(\bm x |\bm u_i)$ for $\bm u_i \in \mathscr{L}_{\omega}$ is the same, denoted by $P_{\omega}(\bm x)$. We have
\begin{equation}
	\begin{aligned}
		{\rm FER}(\omega) &\leq \sum\limits_{\bm y\in \mathcal{Y}^{m}}P(\bm y|\bm 0)\Bigg[\binom{k}{\omega}\sum_{\bm x\in\mathbb{F}_{2}^{m}} P_{\omega}(\bm x)\frac{(P(\bm y|\bm x))^{s}}{(P(\bm y|\bm 0))^{s}}\Bigg]^\gamma\\	
		&\overset{(*)}{\leq}\exp\bigg(-mE(\rho_\omega,R_\omega)\bigg)\text{,}	
	\end{aligned}
	\label{bound2}
\end{equation}
where the inequality $(*)$ follows  from \eqref{boundnumber}, the proof of Lemma 4 and by denoting
\begin{equation}
		\setlength{\abovedisplayskip}{2pt}
	\setlength{\belowdisplayskip}{2pt}
R_\omega=\frac{1}{m}\log{\binom{k}{\omega}}.
\end{equation}

\par For $W_H(\bm u_i)\geq T$, we have
\begin{equation}
	\begin{aligned}
		{\rm Pr}&\{P(\bm y|\bm u_i{\rm {\bf G}})\geq P(\bm y|\bm 0)\}\leq \frac{{\rm{\bf E}}[(P(\bm y|\bm u_i {\rm {\bf G}}))^s]}{(P(\bm y|\bm 0))^s}\\
		&=\sum\limits_{ \bm x\in \mathbb{F}_2^m}P(\bm x|\bm u_i)\frac{(P(\bm y|\bm u_i {\rm {\bf G}}))^s}{(P(\bm y|\bm 0))^s}\\
		&\leq\Bigg[\frac{1+(1-2\rho)^T}{2}\Bigg]^m\sum\limits_{ \bm x\in \mathbb{F}_2^m}\frac{(P(\bm y|\bm u_i {\rm {\bf G}}))^s}{(P(\bm y|\bm 0))^s},
	\end{aligned}
\end{equation}
where the inequality follows from Lemma 2. Thus,  we have

\begin{equation}
\begin{aligned}	
	&\sum\limits_{\bm y\in \mathcal{Y}^m}P(\bm y|\bm 0)\Bigg(\sum\limits_{\bm u_i: W_H(\bm u_i)\geq T}{\rm Pr}\{P(\bm y|\bm u_i{\rm {\bf G}})\geq P(\bm y|\bm 0)\}\Bigg)^{\gamma}\\	
	&\leq \sum\limits_{\bm y\in \mathcal{Y}^m}P(\bm y|\bm 0)\Bigg\{\sum\limits_{\omega>T}|\mathscr{L}_{\omega}|
    \cdot\Bigg[\frac{1+(1-2\rho)^T}{2}\Bigg]^{m}\sum\limits_{\bm x \in \mathbb{F}_2^m} \frac{(P(\bm y|\bm x))^{s}}{(P(\bm y|\bm 0))^{s}}\Bigg\}^{\gamma}\\
	&\overset{(*)}{\leq}\exp\Big[{k\gamma (H(U|V)+2\epsilon)}\Big]\Big[1+(1-2\rho)^T\Big]^{m\gamma}\cdot \Bigg[\sum\limits_{ y_i\in \mathcal{Y}}(P(y_i|0))^{1-s\gamma}\Bigg(\sum\limits_{ x_i\in \mathbb{F}_2}\frac{1}{2}\cdot{(P( y_{i}|x_{i}))^{s}}\Bigg)^{\gamma}\Bigg]^{m}\\
	&\overset{(**)}{\leq} \exp\Big({-mE\Big(\frac{1}{2},{R}_T\Big)}\Big)\text{,}
\end{aligned}
\label{latterbound}
\end{equation}
where the inequality $(*)$ follows from Lemma 3 and the memoryless BIOS channel assumption, and the inequality $(**)$ follows the proof of Lemma 4 and by denoting

\begin{equation}
	{R}_T= \log[1+(1-2\rho)^T]+ \frac{k}{m} (H(U|V)+2\epsilon)\text{.}
\end{equation}

\par Thus, we have
\begin{equation}
	\begin{aligned}
		{\rm{Pr}}\{{\rm error}|\bm u^k\}&\leq \epsilon+\sum\limits_{\omega=1}^{T}\exp\bigg(-mE(\rho
		_{\omega},R_\omega)\bigg)\\
		&\qquad+\exp\Big({-mE\Big(\frac{1}{2},{R}_T\Big)}\Big)\text{.}
		\label{result}
	\end{aligned}
\end{equation}

For the information length $k$, we may choose the parity-check length $m = \lfloor k/r \rfloor $. Now letting $k \rightarrow \infty$ and $T \rightarrow \infty$, we have ${R}_T \rightarrow r(H(U|V)+2\epsilon) $ since $\rho \leq 1/2$ and $k/m \rightarrow r$.
We may choose sufficiently large $k$~(hence $m$) and $T$, such that ${R}_T < I(X; Y ) \leq I_0(1/2) $ since $r < I(X; Y) / H(U|V)$ by assumption. We have from Lemma 4 that $E(1/2, {R}_T) > 0 $ and the third term in the right hand side~(RHS) of the inequality~\eqref{result} can be made not greater than $\epsilon$.
By fixing $T$ and for $\omega < T$, since $m$ increases linearly with $k$ but $\log{\binom{k}{\omega}}$ increases only logarithmically with $k$, we have $R_\omega \rightarrow 0 $ as $k \rightarrow \infty$.
Since $I (\rho_\omega)> 0$, we have $E(\rho_{\omega},R_\omega) > 0$ for sufficiently large $k$ and hence $m$, implying that the second term in the RHS of~\eqref{result} can also be made not greater than $\epsilon$.
Now we have
\begin{equation}
{\rm Pr}\{{\rm error}|\bm u^k\} \leq 3\epsilon.
\end{equation}
Therefore, ${\rm Pr}\{{\rm error} \} = \sum_{\bm u^k \in \mathbb{F}_2^{k} } {\rm Pr}\{{\rm error}|\bm u^k\} \leq 3 \epsilon$.
\hfill $\square$



\section{Simulation Results for JSCC}
\label{sec4}
\par  We have proved in Theorem~2 that, like the totally random linear codes, BGMCs can achieve the theoretical limits for binary sources and BIOS channels. For $\rho\ll1/2$, the BGMCs are low density generator~(LDGM) codes. However, even with this setup, the maximum-likelihood~(ML) decoding is not implementable, so we turn to a very special class of convolutional LDGM codes~\cite{2020SCLDGM} in this section.
\subsection{Convolutional LDGM Codes}
\begin{figure}[tbp]
	\centering
	\includegraphics[width=0.7\textwidth]{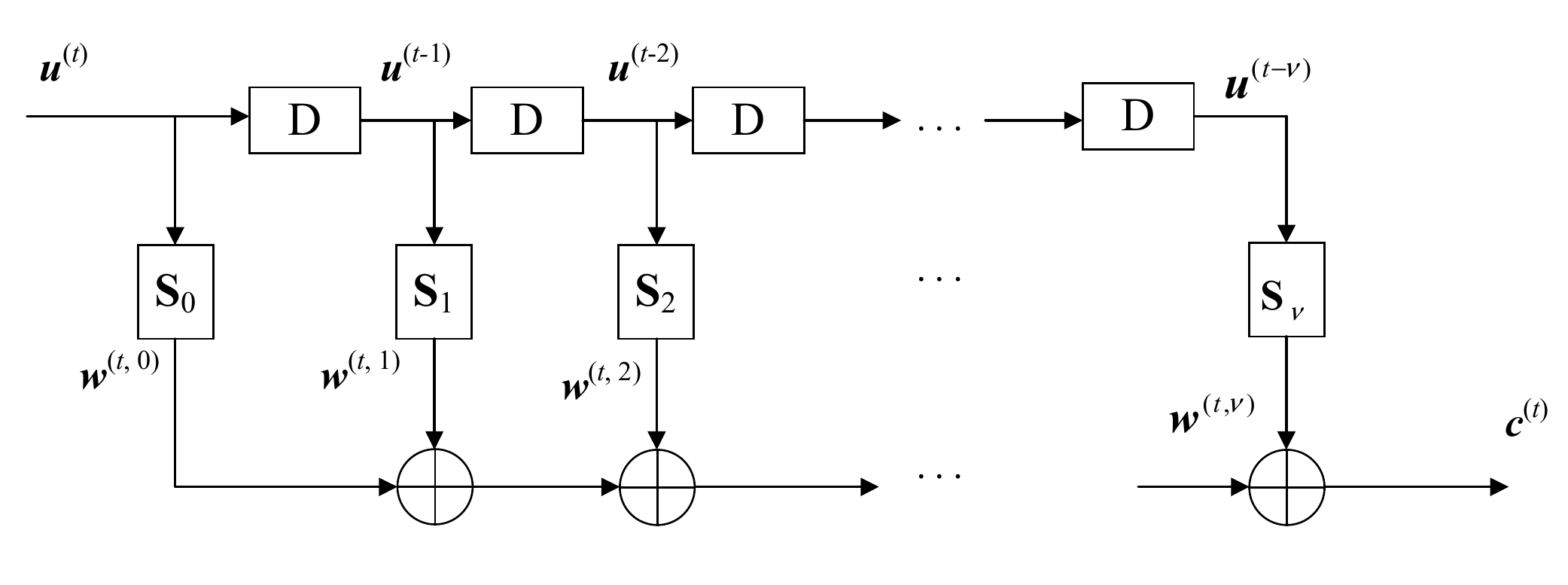}
	\caption{Encoding structure of a convolutional LDGM code with memory $\nu$.}
	\label{SCLDGM}
\end{figure}

  The generator matrix of a time-invariant convolutional LDGM code can be written as  
\begin{equation*}
	\mathbf{G} = \left[
	\begin{array}{cccccccc}
		\mathbf{S}_{0} & \mathbf{S}_{1} &  \cdots  & \mathbf{S}_{\nu-1}  & \mathbf{S}_{\nu}   &     & \\
		& \mathbf{S}_{0} & \mathbf{S}_{1} &  \cdots  & \mathbf{S}_{\nu-1}  & \mathbf{S}_{\nu} &\\ 
		&         & \ddots  & \ddots    & \ddots    & \ddots  & \ddots \\
	\end{array}
	\right]
\end{equation*}
where $\textbf{S}_i$ is a random matrix of size $k\times m$ with each column drawn independently and uniformly from  the collection of all binary column vectors of weight 1. That is to say, the column weights of $\textbf{S}_i$ are confined to be one and hence the generator matrix is very sparse for large $k$. The corresponding encoder as shown in Fig.~\ref{SCLDGM} is almost the same as that for Ensemble 3 presented in~\cite{Ma2016Coding} and the slight difference lies in that the matrices $\textbf{S}_i$ are time-invariant without zero columns and the systematic bits are totally punctured. It is worth pointing out that, in the case when $m(\nu+1)=k\gamma$ for some positive integer $\gamma$, the $\nu +1$ matrices $\textbf{S}_i$, $0\leq i \leq \nu$, can be constructed by partitioning equally $\gamma$ random permutation matrices $\mathbf{\Pi}_i$, $1 \leq i \leq \gamma$, of order $k$.  That is, $[\mathbf{S}_{0}, \mathbf{S}_{1},  \cdots , \mathbf{S}_{\nu-1}, \mathbf{S}_{\nu}] $ is another form of $ [\mathbf{\Pi}_1, \mathbf{\Pi}_2, \cdots, \mathbf{\Pi}_{\gamma}]$. With this setup, the construction is similar to that presented~\cite{Zhu2022BMSTR} but the encoder structure here is different from that presented in~\cite{Zhu2022BMSTR}.

The input to the encoder is a sequence $\bm u^{(0)},~\bm u^{(1)},\cdots$, where  $\bm u^{(i)}\in \mathbb{F}_2^k$. The encoding algorithm of the convolutional LDGM codes is described  in Algorithm~\ref{encoding_algorithm}~(see Fig.~\ref{SCLDGM} for reference).

\begin{algorithm}
	\caption{Encoding of convolutional LDGM codes}
	\label{encoding_algorithm}
	\begin{itemize}
		\item \textbf{Initialization}:\; For $t<0$, let ${\bm {u}}^{(t)}=\textbf{0}\in \mathbb{F}^{k}_{2}$.
		\item \textbf{Iteration}: For $0\leq t\leq L-1$,
		\begin{itemize}
			\item For $0\leq i\leq \nu$, compute ${\bm{w}}^{(t,i)}={\bm{u}}^{(t-i)}\textbf{S}_\emph{i}\in\mathbb{F}^{m}_{2}$.
			\item Compute ${\bm c}^{(t)}=\sum_{i =0}^{\nu}{\bm w}^{(t,i)}$.
			\item Take $\bm c^{(t)}$ as the encoding output block at time slot $t$.
		\end{itemize}
		\item \textbf{Termination}: For $L\leq t\leq L+m-1,$ set $\bm{u}^{({t})}=\textbf{0}\in\mathbb{F}^{k}_{2}$ and compute $\bm{c}^{(t)}$ following Step \textbf{Iteration}.
	\end{itemize}
	
\end{algorithm}
\par For the decoding, we use an iterative sliding window  decoding algorithm, which is slightly different from but similar to the decoding algorithm as described in~\cite[Algorithm~3]{2020SCLDGM}. The  difference lies in that the systematic digits are totally erased but initialized according to the source distribution.
\par
\begin{figure}[tbp]
	\centering
	\includegraphics[width=0.7\textwidth]{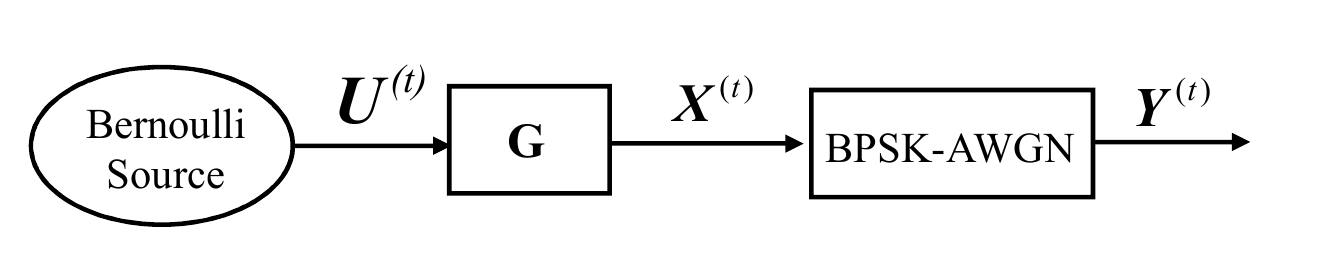}
	\caption{System model with convolutional LDGM codes for JSCC.}
	\label{JSCC}
\end{figure}

 It has been shown that the above convolutional LDGM codes perform well as source codes~\cite{Zhu2021Compression} and as channel codes~\cite{2020SCLDGM}. We will show that the convolutional LDGM codes are also good as JSCCs. As shown in Fig.~\ref{JSCC}, we consider transmission of a Bernoulli source with binary phase shift keying~(BPSK) signalling over an additive white Gaussian noise~(AWGN) channels. The binary sparse source delivers  $\bm U^{(t)}\in \mathbb{F}_2^{k}$ at time slot $t$, which is an i.i.d Bernoulli source with success probability $\theta\triangleq\text{\rm Pr}\{U^{(t)}_i=1\}$. The entropy of this source is given by
\begin{equation}
	H(U)=-\theta\log\theta-(1-\theta)\log(1-\theta)\triangleq H(\theta). 
\end{equation} 
 The source block  $\bm U^{(t)}$ is encoded by the convolutional LDGM code with the generator matrix $\textbf{G}$ into $\bm X^{(t)}\in \mathbb{F}_2^{m}$. Then $\bm{X}^{(t)}$ is  modulated using BPSK signalling with 0 and 1 mapped to $+1$ and $-1$, respectively, and transmitted over an AWGN channel characterized by the noise power $\sigma^2$, resulting in $\bm Y^{(t)}\in \mathbb{R}^m$. The task of the receiver is to get the estimation of the source $\hat{\bm U}^{(t)}\in \mathbb{F}_2^k$ from $\bm Y^{(t)},\cdots,\bm Y^{(t+d)}$ using the iterative sliding window decoding algorithm.  

\begin{figure}[tbp]
	\centering
	\includegraphics[width=0.7\textwidth]{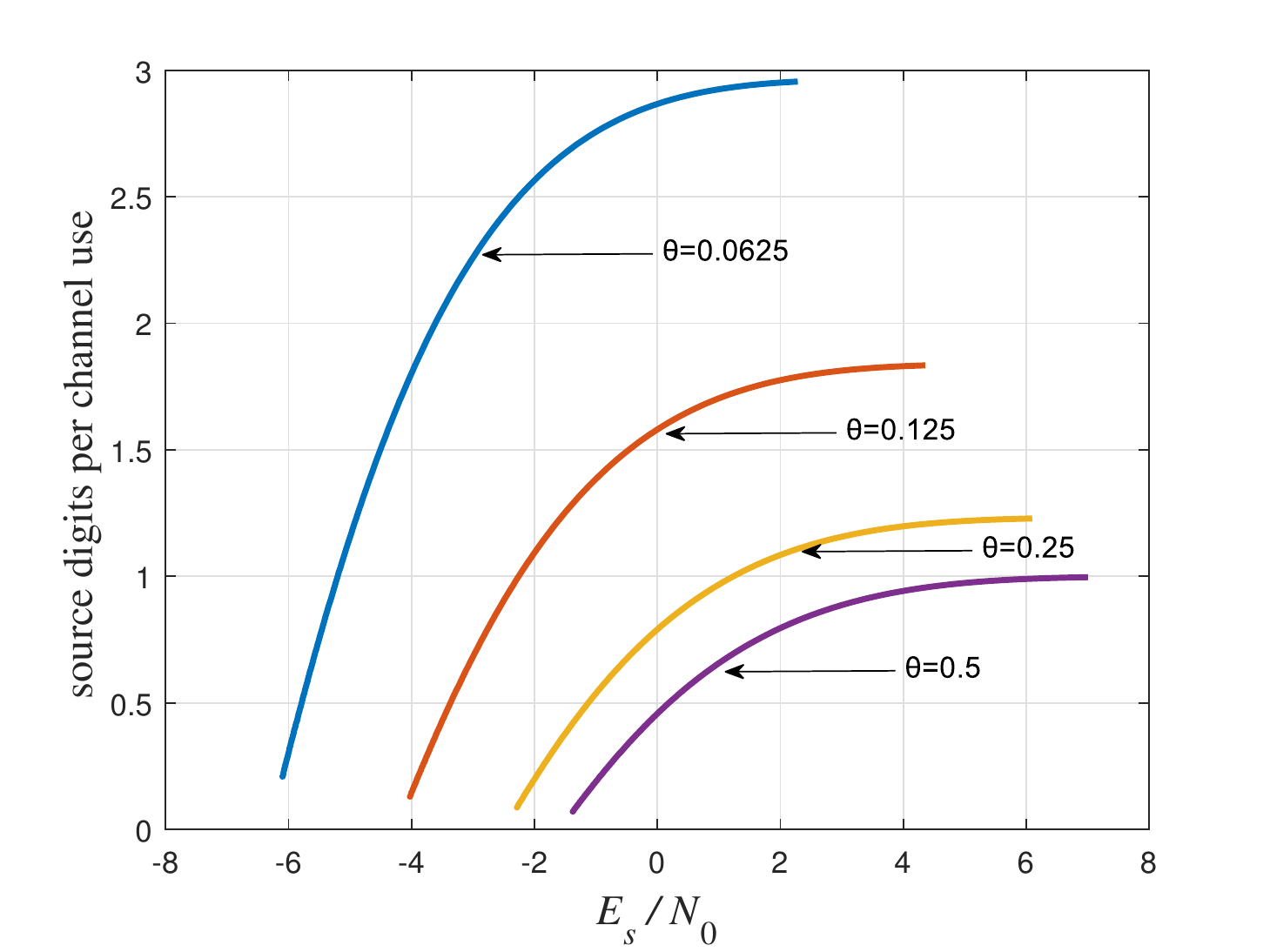}
	\caption{System capacity with different $\theta$. Notice that the ordinate is measured in source digits per channel use  instead of the commonly-adopted bits per channel use. Hence, the value of the ordinate can be greater than one for sparse sources. The abscissa is measured by $E_s/N_0$ where  $E_s$ and $N_0$ refer to the energy per source digit and  the noise power spectral density, respectively. }
	\label{systemCapacity}
\end{figure}
\par We define the code rate of the JSCC system as $R=k/m$ measured in source digits per channel use. From Theorem 1,  the supremum of the code rate $R$ can be $C/H(\theta)$, which is measured in source digits per channel use and referred to as the \emph{system capacity}  to distinguish from the conventional channel capacity. Fig.~\ref{systemCapacity} shows the system capacity with different $\theta$, where we can see that the sparser the source is, the more source digits can be carried on average by one channel use. In this paper, $E_s$ refers to the energy per source digit~(rather than nonredundant information bit), the same as that in~\cite{Lau2021}, and  $N_0=2\sigma^2$ refers to the noise power spectral density and $\sigma^2$ is the variance of noise. Thus, the limits of $E_s/N_0$ in Fig.~\ref{systemCapacity} can be less than $-1.59~{\rm dB}$, which is the asymptotic limit  for reliable transmission when the energy is measured per non-redundant information bit, denoted by $E_b$. Actually, we have
\begin{equation*}
	\frac{E_b}{N_0}=\frac{E_s}{N_0}-10\log_{10}(H(\theta)).
\end{equation*}
Specially, when the source is uniform, i.e., $\theta=0.5$, we have $E_s/N_0=E_b/N_0$.

\subsection{Performance Bound }
\par  We define the BER which represents the source digit error rate and is referred to as the average Hamming distortion for biased sources,  as $\textbf{E}[W_H(\hat{\bm U}^{(t)}+\bm U^{(t)})]/k$, where $\textbf{E}[\cdot]$ denotes the expectation of the random variable, $W_H(\cdot)$ denotes the Hamming weight function.  We first present a lower bound on the BER of the linear block codes, which is also applicable to the convolutional LDGM codes. 
 \par \emph{Theorem 3:} For the BPSK-AWGN channel and sparse source, the BER of a linear block code with generator matrix $\textbf{G}$ of size $k\times m$ can be lower bounded by
 
\begin{equation}
	\begin{aligned}
		{\rm BER}&\geq \frac{1}{k}\sum\limits_{i=0}^{k-1}P_{\theta,W}(\theta,\omega_i),
	\end{aligned}
\end{equation}
where  $P_{\theta,W}(\theta,\omega_i)$ is the  lower bound on the bit error probability for the $i$-th bit, given by
\begin{equation}
	\begin{aligned}
	P_{\theta,W}(\theta,\omega_i)=&(1-\theta) Q(\frac{\sqrt{\omega_i}}{\sigma}+\frac{\sigma}{2\sqrt{\omega_i}}\ln\frac{1-\theta}{\theta})+\theta \cdot Q(\frac{\sqrt{\omega_i}}{\sigma}-\frac{\sigma}{2\sqrt{\omega_i}}\ln\frac{1-\theta}{\theta}).
	\end{aligned}
\end{equation}
Here, $\theta$ is the success probability of the Bernoulli source, $\omega_i$ is the Hamming weight of the $i$-th row of $\textbf{G}$, $\sigma^2$ is the variance of the noise and $Q(x)$ is the tail probability that the normalized Gaussian random variable takes a value not less than $x$.
\par \emph{Proof:} Let $\omega_i$ be the Hamming weight of the $i$-th row of $\textbf{G}$. The error probability of the $i$-th bit $U_i$ is lower bounded by that with the genie-aided decoder\cite{Ma2015BMST}, which assumes that all but $U_i$ is known. Therefore, the lower bound is equal to the performance of repetition code with length $\omega_i$.
 Without loss of generality, we assume that the receiving sequence corresponding to the repetition code is $\bm y=(y_0,\cdots,y_{\omega_i-1})$. Upon receiving $\bm y$, the optimal decision with log-likelihood ratio~(LLR)  for the repetition code is given by
\begin{equation}
	\hat{U_i}=\left\{\begin{aligned}
 		&0,\qquad \ln(\frac{1-\theta}{\theta})\sum\limits_{j=0}^{\omega_i-1}\frac{2y_j}{\sigma^2}>0\\
		&1,\qquad \ln(\frac{1-\theta}{\theta})\sum\limits_{j=0}^{\omega_i-1}\frac{2y_j}{\sigma^2}<0
	\end{aligned}\right.\text{.}
	\label{flipping}
\end{equation}
 For the optimal decision,  the error probability for bit 0 is $Q(\sqrt{\omega_i}/{\sigma}+\sigma/{(2\sqrt{\omega_i})}\ln[(1-\theta)/{\theta}])$ and for bit 1 is $Q(\sqrt{\omega_i}/{\sigma}-\sigma/{(2\sqrt{\omega_i})}\ln[(1-\theta)/{\theta}])$. Hence,  the error probability for the repetition code with length $\omega_i$ is
\begin{equation}
	\begin{aligned}
	P_{\theta,W}(\theta,\omega_i)=& (1-\theta) Q(\frac{\sqrt{\omega_i}}{\sigma}+\frac{\sigma}{2\sqrt{\omega_i}}\ln\frac{1-\theta}{\theta})+\theta \cdot Q(\frac{\sqrt{\omega_i}}{\sigma}-\frac{\sigma}{2\sqrt{\omega_i}}\ln\frac{1-\theta}{\theta}).
	\end{aligned}
\end{equation}                
We can get the  result in the theorem by averaging the lower bound on BER of $k$ bits.
\hfill $\square$
\par \emph{Remark:} For $\theta=1/2$, the lower bound on BER for the convolutional LDGM codes is reduced to
\begin{equation}
	{\rm BER}\geq \frac{1}{k}\sum\limits_{i=0}^{k}Q(\frac{\sqrt{\omega_i}}{\sigma}),
\end{equation}  
which is the same as the result derived in \cite[Theorem 2]{2020SCLDGM}.
\par We see that the limit performance is dominated by the row weights, which are in turn closely related to the encoding memory $\nu$. This is confirmed by the following numerical examples, which also show the flexibility  and universality of the construction. Particularly, the convolutional LDGM codes are definitely applicable to the case of the uniform source and the case of the noiseless channel, as already demonstrated in~\cite{Wang2022} and~\cite{Zhu2021Compression}. 
\subsection{Simulation results}

\par
\begin{figure}[tbp]
	\centering
	\includegraphics[width=0.65\textwidth]{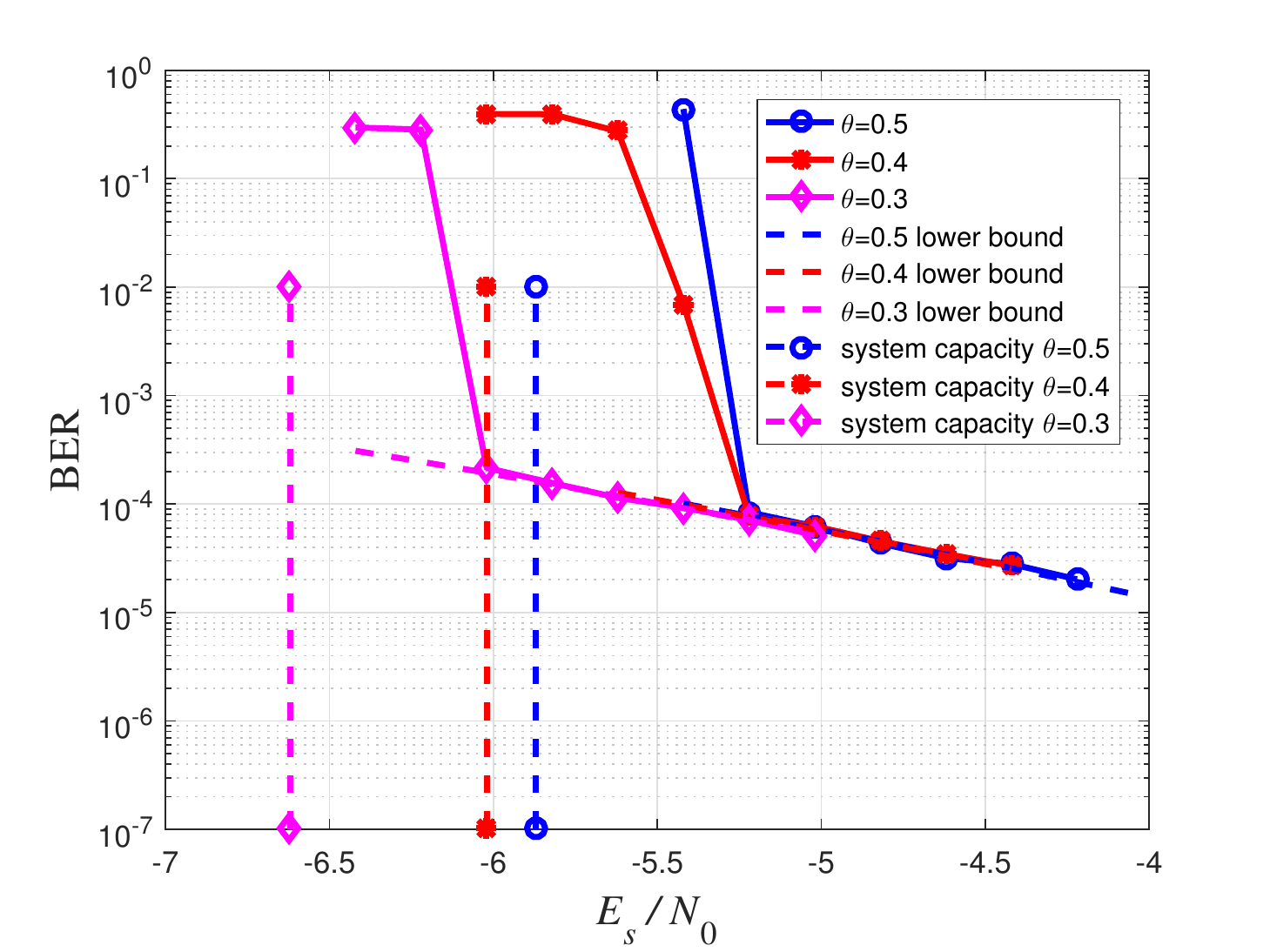}
	\caption{The BER performance of convolutional LDGM codes with $k=1024,m=2048,\nu=7$ in Example 1. The parameter $\theta$ is specified in the legends. The corresponding lower bounds and system capacities are also plotted.}
	\label{m=2048}
\end{figure}

\begin{figure}[htbp]
	\centering
	\includegraphics[width=0.65\textwidth]{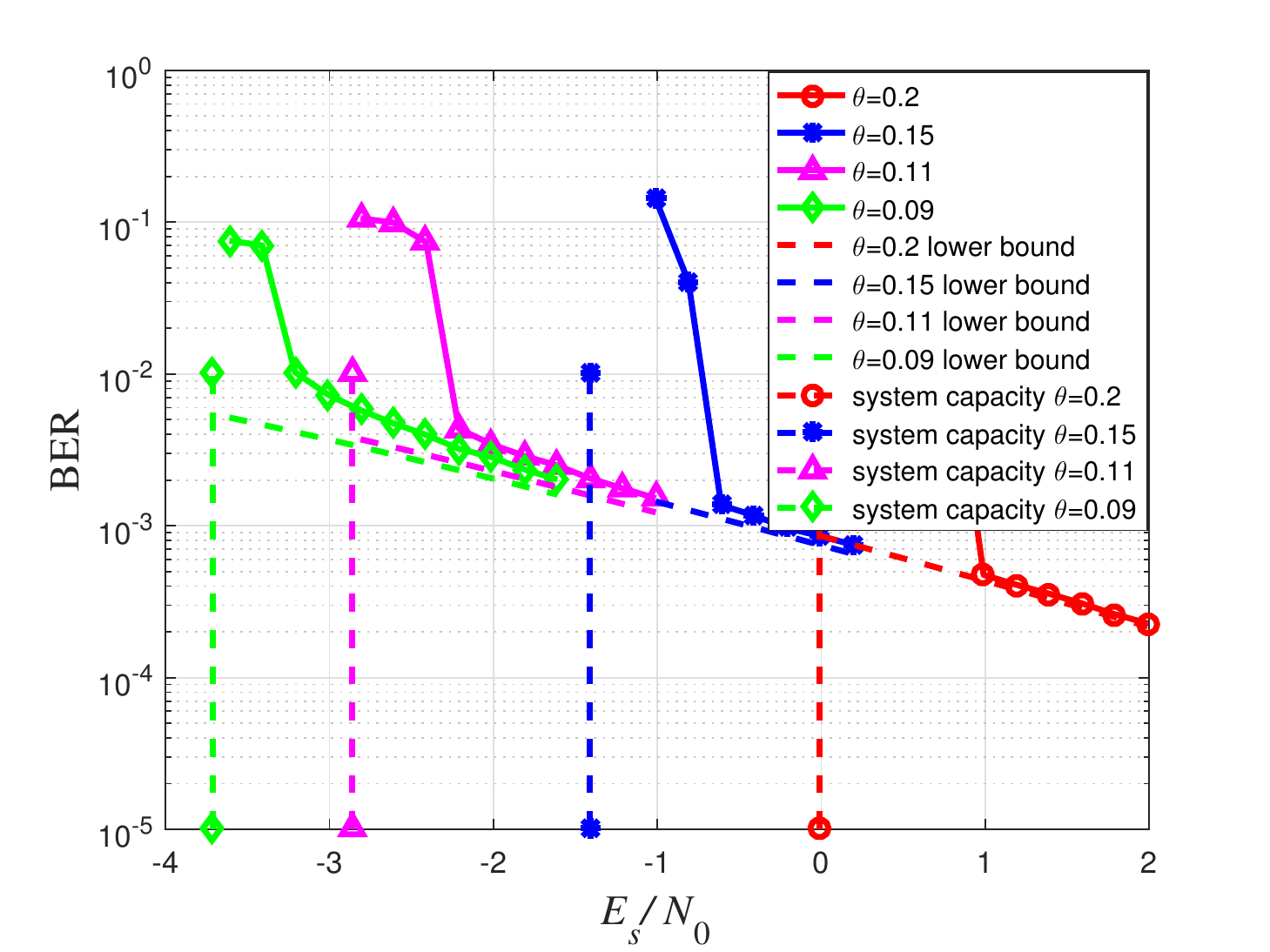}
	\caption{The BER performance of convolutional LDGM codes with $k=1024,m=1024$, $\nu=7$ in Example 1. The parameter $\theta$ is specified in the legends. The corresponding lower bounds and system capacities are also plotted.}
	\label{m=1024}
\end{figure}

\begin{figure}[htbp]
	\centering
	\includegraphics[width=0.65\textwidth]{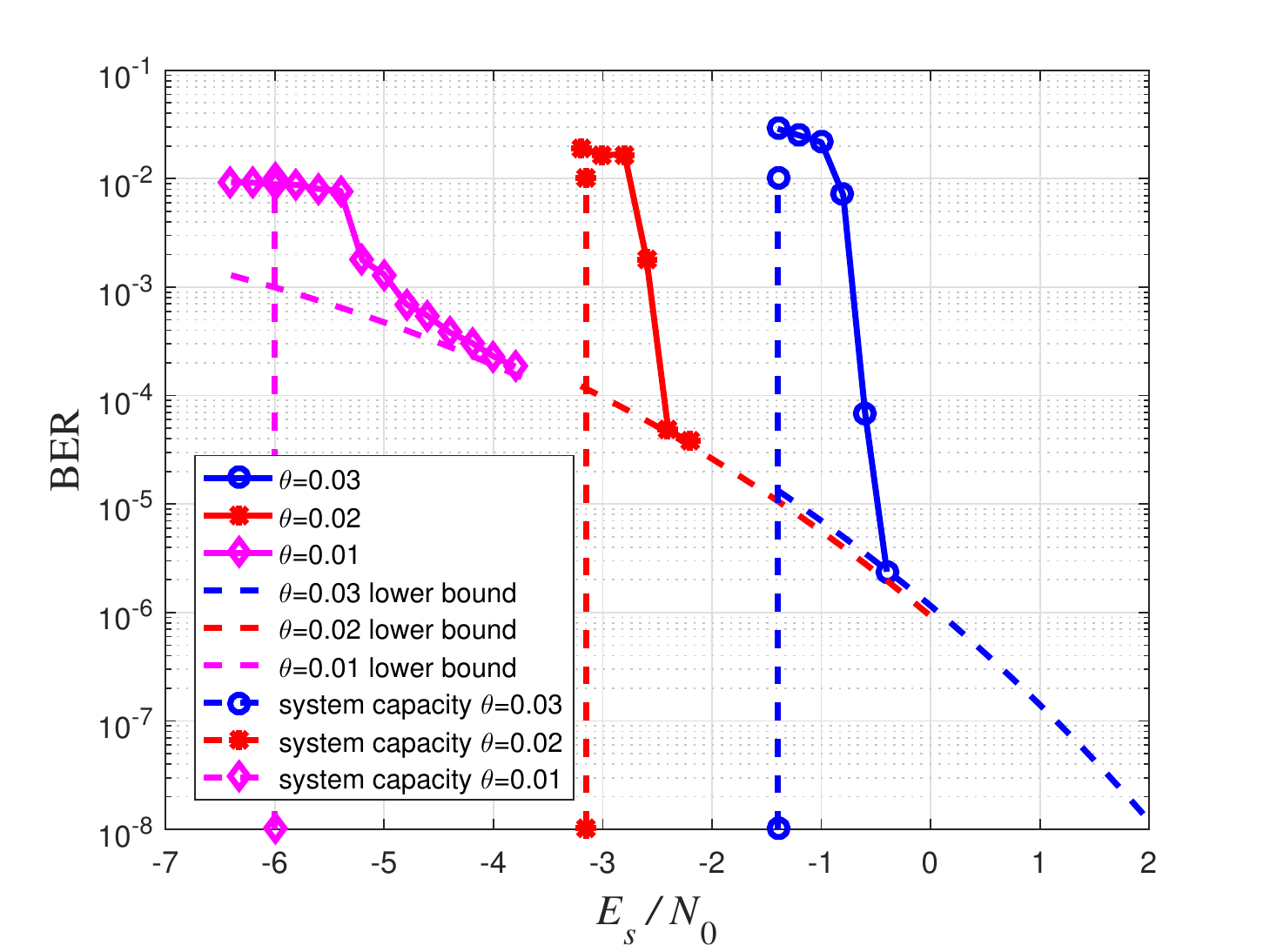}
	\caption{The BER performance of convolutional LDGM codes with $k=1024,m=512,\nu=50$ in Example 1. The parameter $\theta$ is specified in the legends. The corresponding lower bounds and system capacities are also plotted.}
	\label{m=512}
\end{figure}

 \emph{Example 1~(Fixed $k,m$ and $\nu$, Increasing $\theta$):}~In this example, we fix  $k=1024,~m=2048,~\nu=7$ in Fig.~\ref{m=2048},  $k=1024,~m=1024,~\nu=7$ in Fig.~\ref{m=1024},  $k=1024,~m=512,~\nu=50$ in Fig.~\ref{m=512} and change the parameter $\theta$. The simulation results are shown in these figures, in which the system capacities and corresponding lower bounds for different $\theta$ are also plotted. From the figures, we can see that  $\theta$ has little influence on the lower bounds for convolutional LDGM codes. Besides, the waterfall region for convolutional LDGM codes are about one dB away from the system capacity.

\par
\begin{figure}[tbp]
	\centering
	\includegraphics[width=0.65\textwidth]{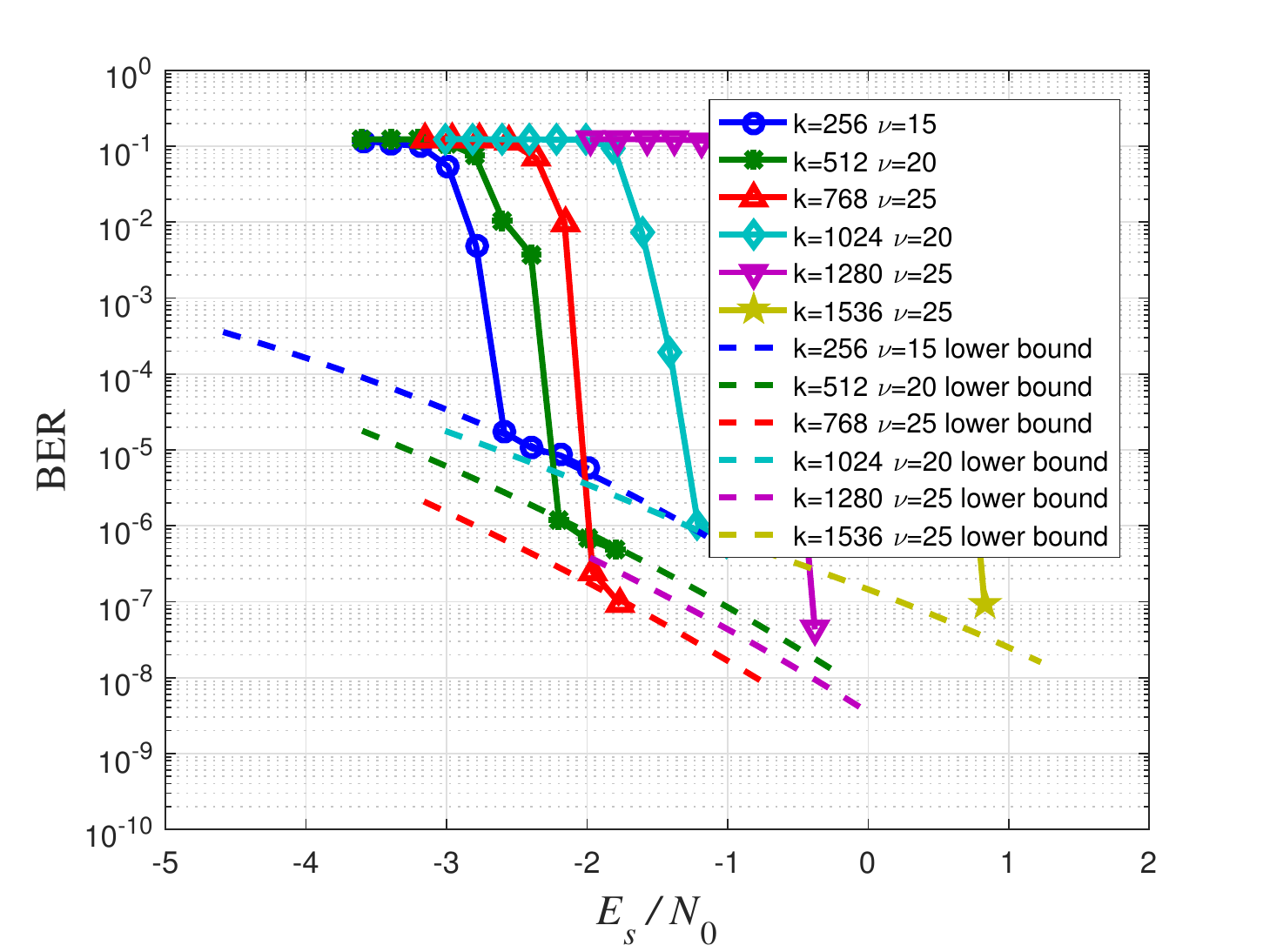}
	\caption{The BER performance of convolutional LDGM codes with  $m=1024,\theta=0.125$  in Example 2. The parameter $k$ and $\nu$ are specified in the legends. The code rates for the convolutional LDGM codes are $1/4,~1/2,~3/4,~1,~5/4$ and $3/2$. The corresponding lower bounds are also plotted.}
	\label{differentK}
\end{figure}

\begin{figure}[tbp]
	\centering
	\includegraphics[width=0.65\textwidth]{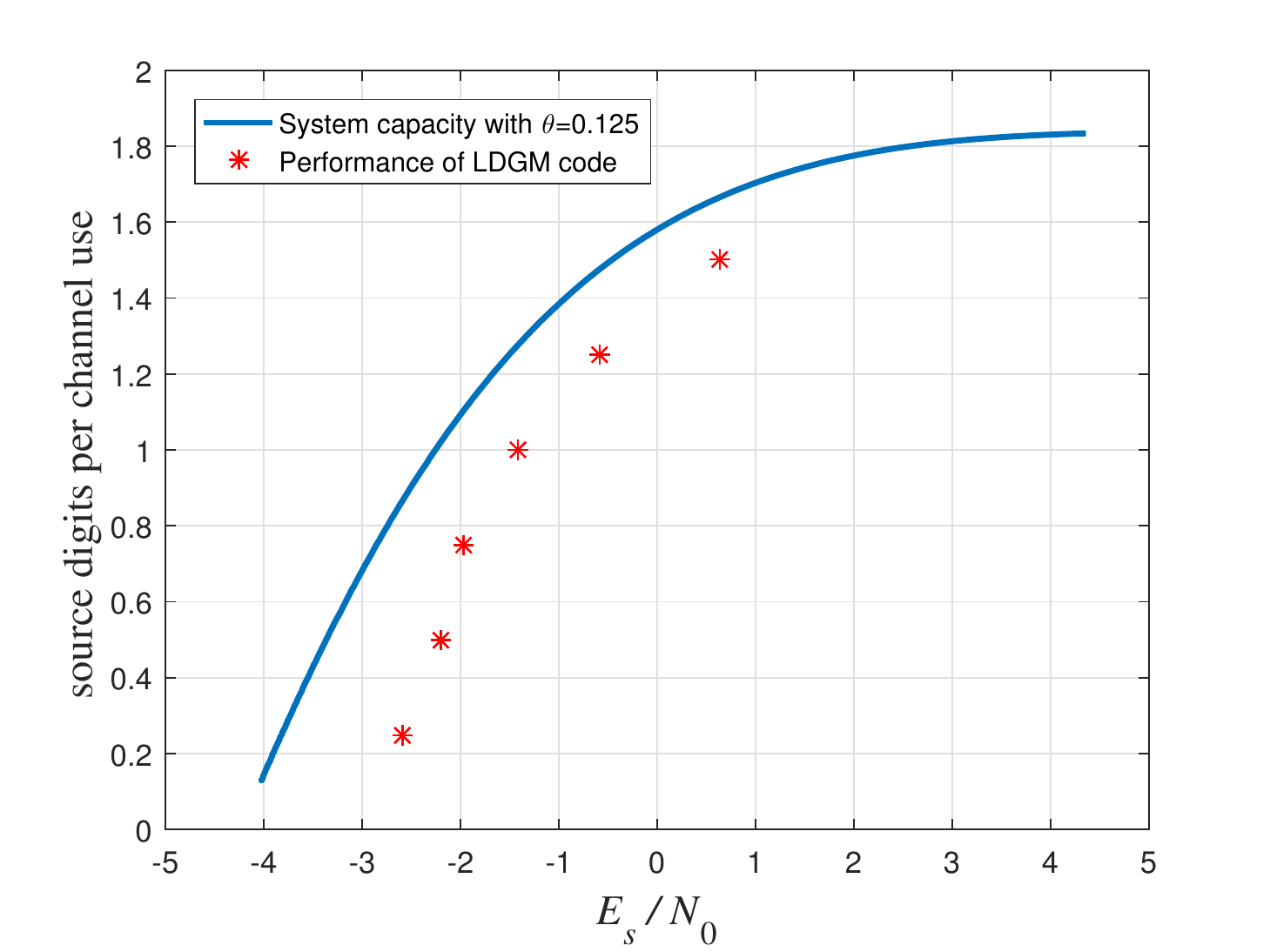}
	\caption{The required SNR for the convolutional  LDGM codes with $m=1024$ to achieve the BER of $10^{-5}$ with BPSK signalling over AWGN channels.}
	\label{capacity_performance}
\end{figure}

 \emph{Example 2~(Fixed $m$ and $\theta$,  Changing $k$ and $\nu$):}~In this example, we fix $m=1024$ and $\theta=0.125$. The BER performance with different values of $k$ and $\nu$ is shown in Fig.~\ref{differentK} and the code rates for the convolutional LDGM codes are $1/4,~1/2~,3/4,~1,~5/4$ and~$3/2$. The corresponding lower bounds are also plotted. From the  simulation results, we
can see that the BER performance of the convolutional LDGM codes match well with the respective  lower bounds in the low BER region for all considered code rates.  We can also observe that the convolutional LDGM codes  achieve the BER of $10^{-5}$ around one dB away from the system capacity for all considered code rates, as shown in Fig.~\ref{capacity_performance}. 

\par
\begin{figure}[tbp]
	\centering
	\includegraphics[width=0.65\textwidth]{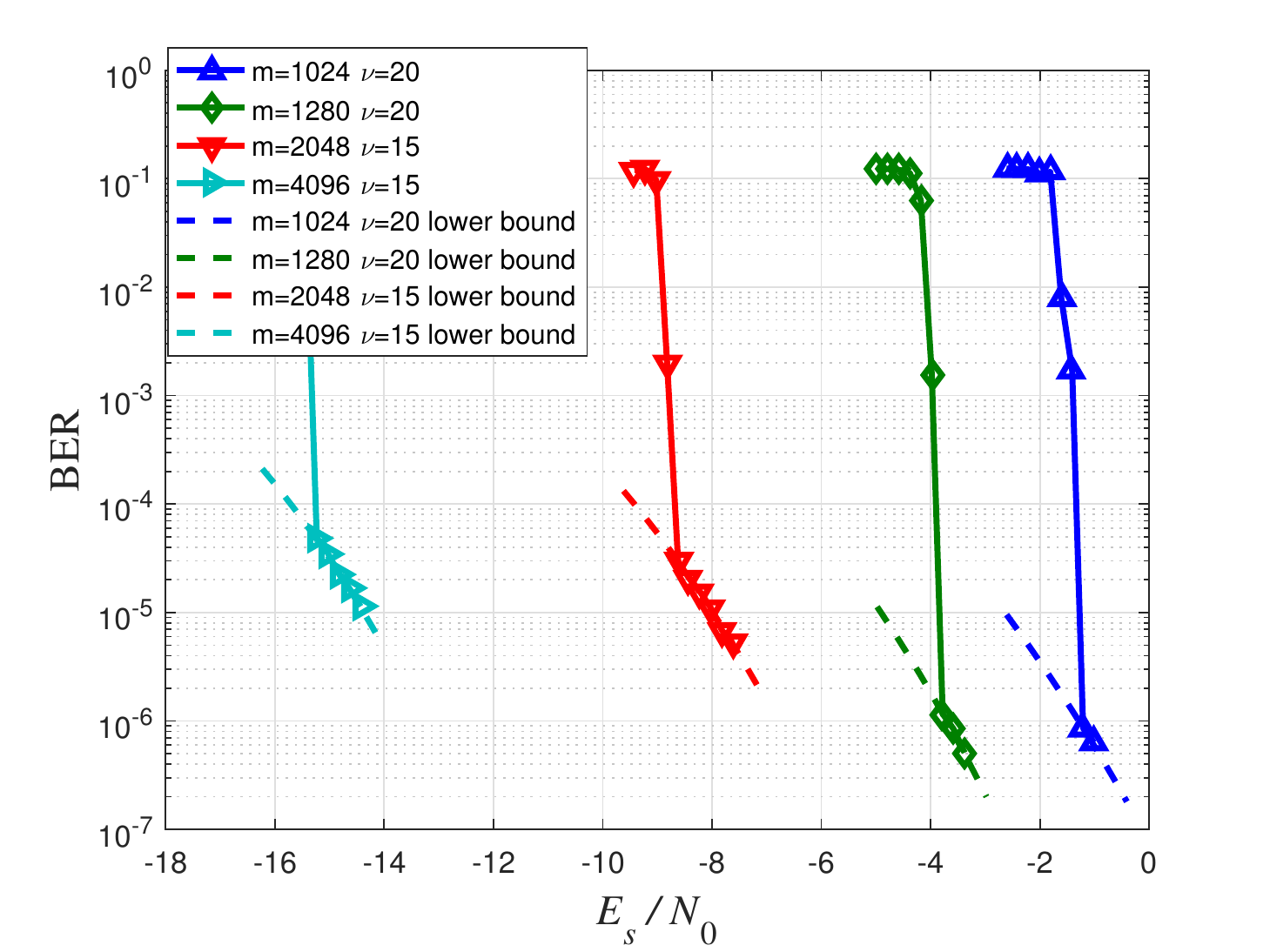}
	\caption{The BER performance of convolutional LDGM codes with  $k=1024,\theta=0.125$  in Example 3. The parameter $m$ and $\nu$ are specified in the legends. The code rates for the convolutional LDGM codes are $1,~4/5,~1/2$ and $1/4$. The corresponding lower bounds are also plotted.}
	\label{fixk}
\end{figure}

 \emph{Example 3~(Fixed $k$ and $\theta$, Changing $m$ and $\nu$):}~In this example, we fix $k=1024$ and $\theta=0.125$. The BER performance with different values of $m$ and $\nu$  are shown in Fig.~\ref{fixk}. The code rates for the convolutional LDGM codes are $1,~4/5,~1/2$ and $1/4$. The corresponding lower bounds are also plotted. From the simulation results, we
can see that the performance of the convolutional LDGM codes match well with the respective  lower bounds in the low BER region for all considered code rates.  

\par
\begin{figure}[tbp]
	\centering
	\includegraphics[width=0.65\textwidth]{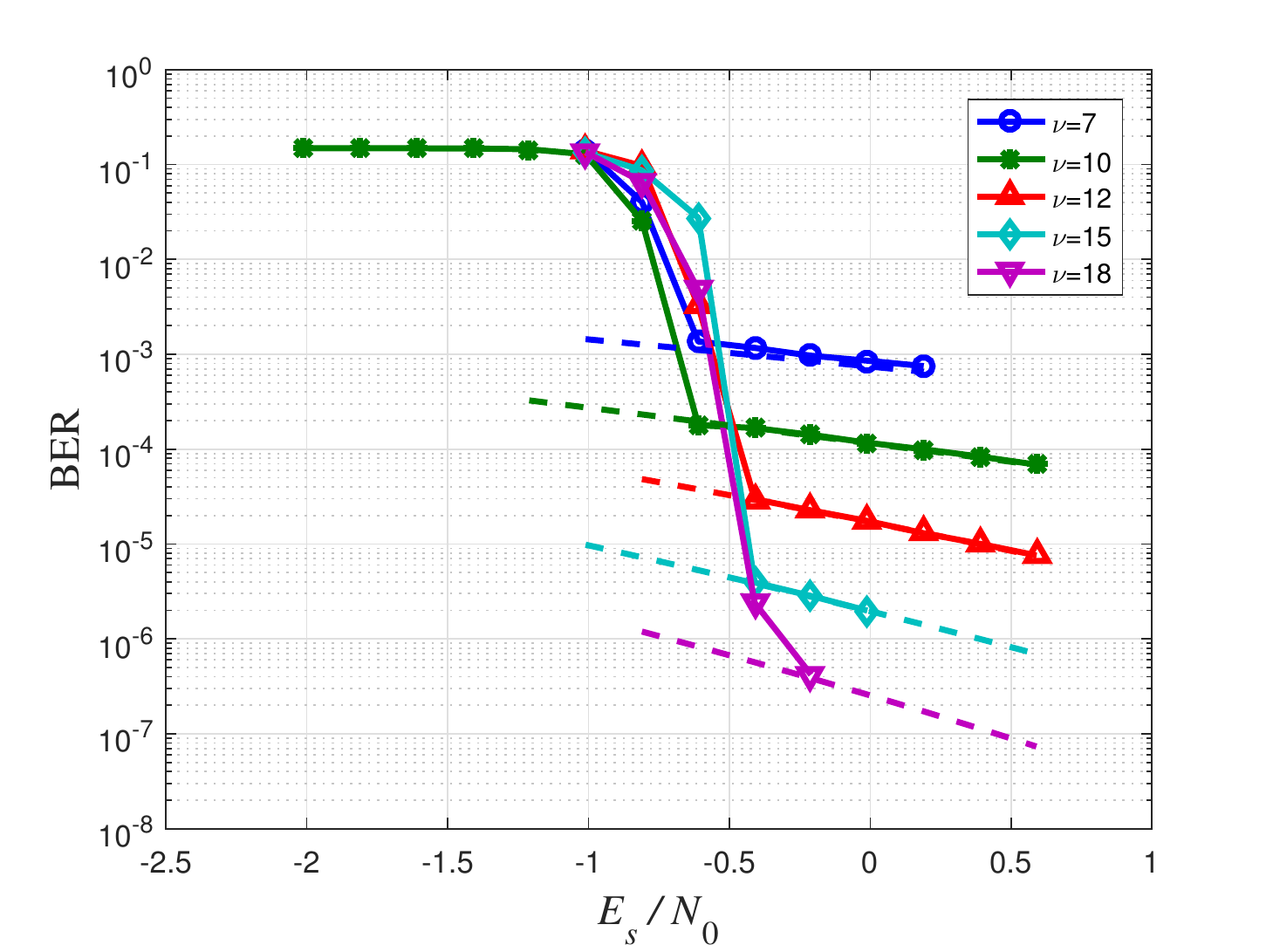}
	\caption{The BER performance of convolutional LDGM codes with $k=1024,m=1024,\theta=0.15$ in Example 4. The parameter $\nu$ is specified in the legends. The corresponding lower bounds  are also plotted.}
	\label{differentnu}
\end{figure}

 \emph{Example 4~(Fixed $k,m$ and $\theta$, Increasing $\nu$):}~In the example, we consider the configuration of  $k=1024, m=1024, \theta=0.15$ for convolutional LDGM codes. The value of encoding memory $\nu$ is changed. The BER simulation results are shown in Fig.~\ref{differentnu}, in which the corresponding lower bounds are also plotted. We can observe from the figure that the error floor can be lowered down by increasing the encoding memory $\nu$.
\par
\begin{figure}[tbp]
	\centering
	\includegraphics[width=0.6\textwidth]{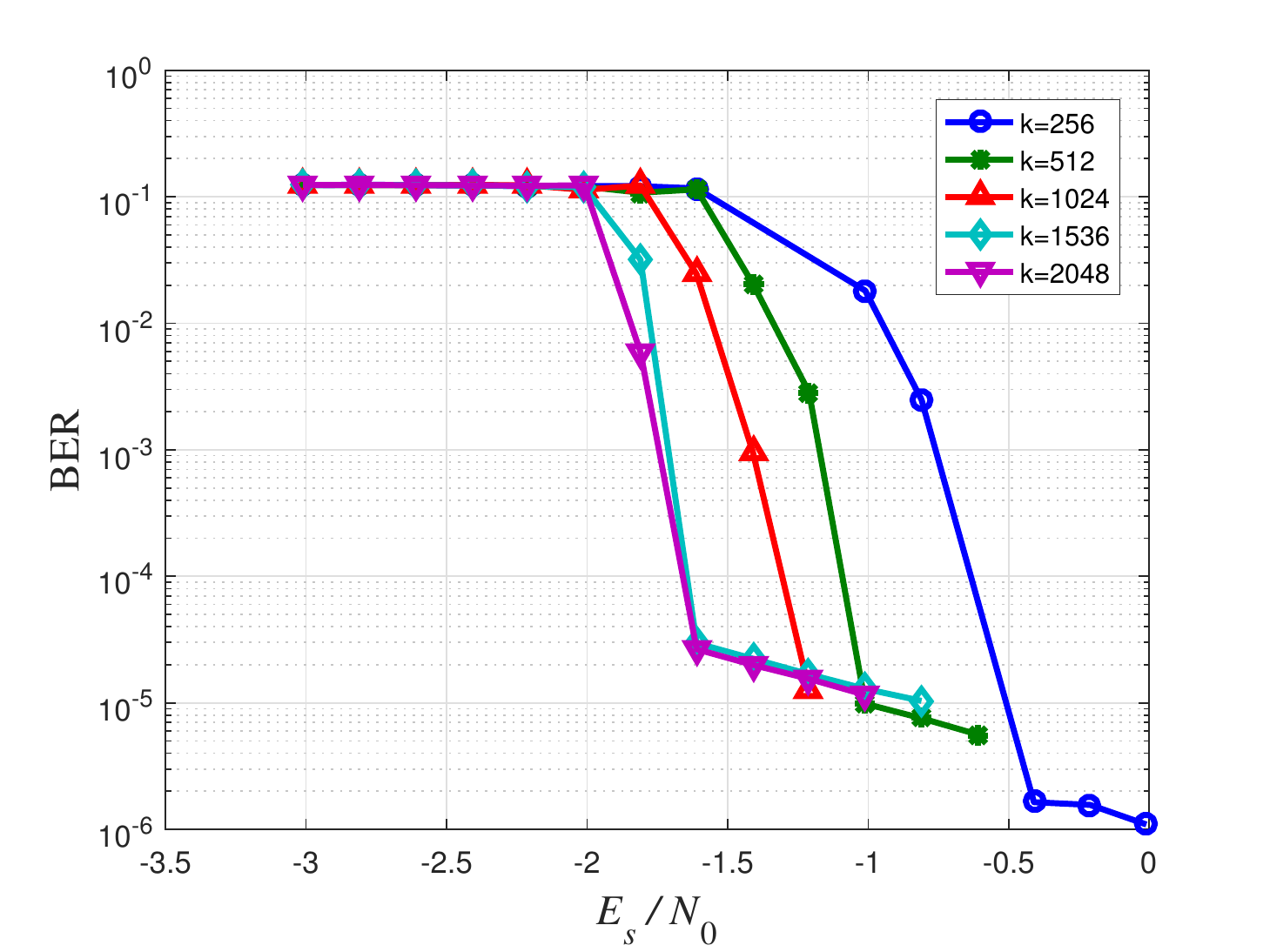}
	\caption{The BER performance of convolutional LDGM codes with  code rate $1$, $\theta=0.125$ and $\nu=15$  in Example 5. The parameter $k$ is specified in the legends.}
	\label{sameRatio}
\end{figure}

 \emph{Example 5~(Fixed code rate $k/m$, $\theta$ and $\nu$, Increasing $k$ and $m$):} In this example, we fix the code rate $R=k/m$ to 1, the value of $\theta$ to 0.125  and encoding memory $\nu$ to $15$. We change the value of $k$ and $m$ in the simulations. The BER simulation results for the convolutional LDGM codes are shown in Fig.~\ref{sameRatio}. We can see from the figure that  the larger value of code length for the convolutional LDGM codes, the better performance we can get in the waterfall region. 
\subsection{Performance Comparison}
\begin{figure}[htbp]
	\centering
	\includegraphics[width=0.65\textwidth]{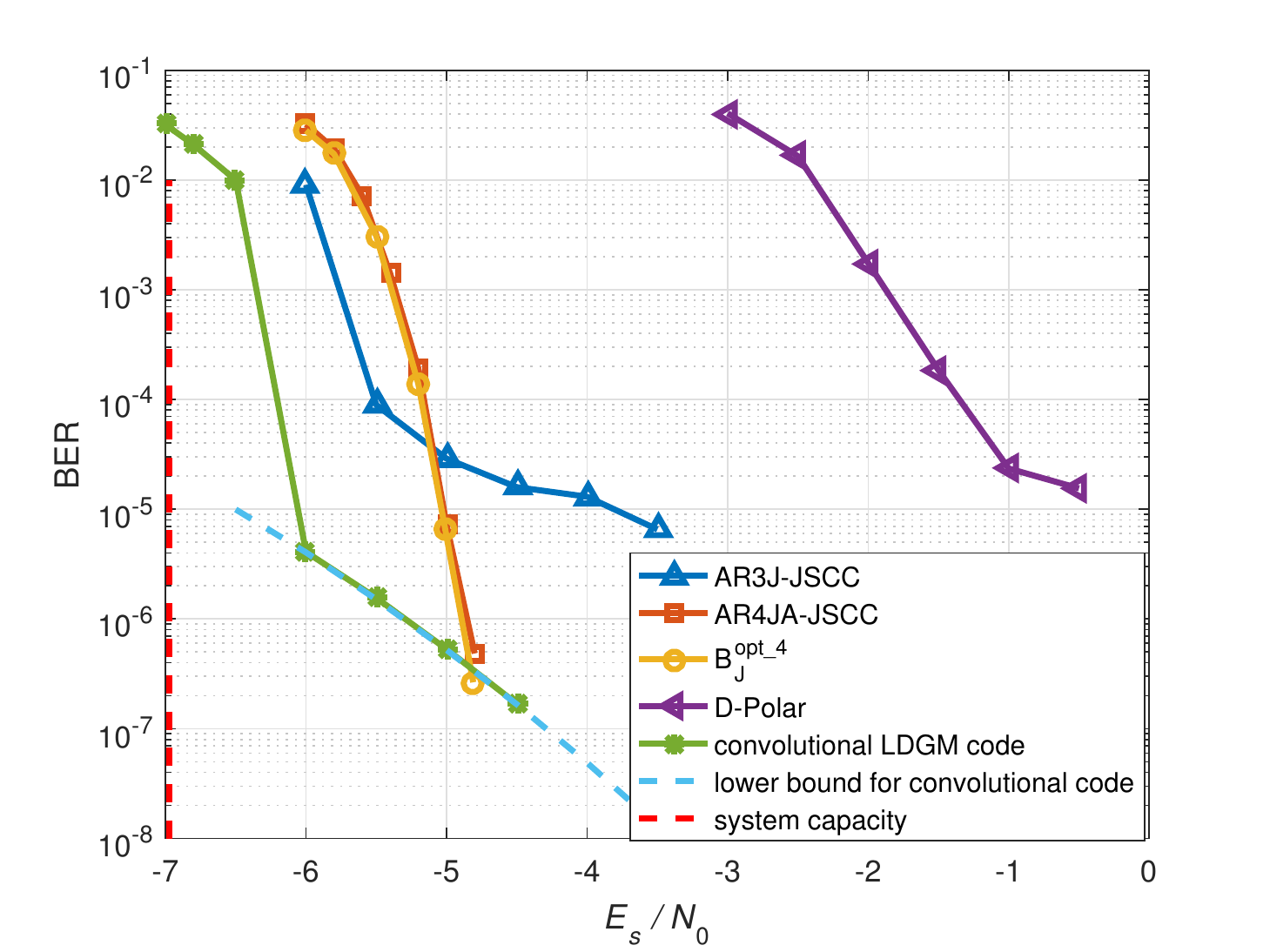}
	\caption{The BER performance of convolutional LDGM codes, AR3A-JSCC codes and AR4A-JSCC codes in~\cite{Lau2021}, ${\textbf B}_J^{opt\_4}$ codes in~\cite{Liu2020} and  D-Polar in~\cite{Dong2021} with $\theta=0.04$. The system capacity for $\theta=0.04$ and the lower bound for convolutional codes are also plotted. }
	\label{comparison}
\end{figure}
\par In Fig.~\ref{comparison}, we present the BER performance for the convolutional LDGM codes with $\theta=0.04,~k=1024,~m=1024,~\nu=40$. We also plot the BER performance of AR3J-JSCC codes and AR4JA-JSCC codes in~\cite{Lau2021}, ~$\textbf{B}_J^{opt\_4}$ codes in~\cite{Liu2020} and  D-Polar in~\cite{Dong2021} for comparison. The system capacity and the lower bound for convolutional LDGM codes are also plotted. The results show that the convolutional LDGM codes have better waterfall region and error floor performance compared with AR3J-JSCC codes and D-Polar codes. Although no error floor down to  BER of $10^{-6}$ has been observed with AR4JA-JSCC codes and $\textbf{B}_J^{opt\_4}$ codes, the convolutional LDGM codes can achieve better performance in the waterfall region, which is about one dB away from the system capacity. Notice that the error floor of convolutional codes can be further lowered down simply by increasing the encoding memory $\nu$, as shown in Fig.~\ref{differentnu}. At ${\rm BER}=10^{-5}$~and~$10^{-6}$, the convolutional LDGM codes have about one~dB and $0.5$ dB gain compared to AR4JA-JSCC codes and $\textbf{B}_J^{opt\_4}$ codes. It is worth to noting that the above comparison is not fair, since the convolutional LDGM codes need large decoding window and have decoding delay of $d=2\nu$. However, the convolutional LDGM codes are stream-oriented and have a fix coding delay of $d$.
\section{Conclusions}
\label{sec5}

In this paper, we have presented a new framework with linear codes for transmission of Bernoulli sources and   proved the coding theorems by deriving partial error exponents.
This new framework allows us to unify the proofs of the lossless source coding theorem, the channel coding theorem and the source-channel coding theorem. We  derive system capacity  for JSCC from the coding theorem and the lower bound for linear codes for performance analysis. A special class of linear codes called convolutional LDGM codes are considered in JSCC scheme for simulations.   The simulation results  show the flexibility of the construction and the predicable performance in error floor region of the convolutional LDGM codes. To lower down the error floor, we can simply increase the encoding memory $\nu$, as shown in Example~4. Under iterative sliding window  decoding algorithm, the convolutional LDGM codes have good waterfall region performance and can achieve about one dB away from the system capacity in the waterfall region for various code rates.

\section{Acknowledgement}
The authors would like to thank Dr. Suihua Cai from Sun Yat-sen University and Prof. Kai Niu from Beijing University of Posts and Telecommunications for their helpful discussions. 

\bibliographystyle{IEEEtran}
\bibliography{bibliofile}


%
%
%
%
%
%
%

\end{document}